\documentstyle[12pt,epsf]{article}
\begin{document}
\title{Stabilisation of Emulsions by Trapped Species}
\author{A. J. Webster$^*$ and M. E. Cates\\ Department of Physics and
Astronomy,\\ University of Edinburgh, JCMB King's Buildings,\\
Mayfield Road, Edinburgh EH9 3JZ, UK.\\}
\maketitle
\date

\begin{abstract}

We consider an emulsion whose droplets contain a trapped species
(insoluble in the continuous phase), and study the emulsion's stability against
coarsening via Lifshitz--Slyozov dynamics (Ostwald Ripening).
Extending an earlier treatment by Kabalnov et al({\em Colloids and
Surfaces}, {\bf 24} (1987), 19-32), we derive a general
condition on the mean initial droplet volume which ensures stability, even
when arbitrary polydispersity is present in both size
and composition of the initial droplets. We distinguish ``nucleated''
coarsening, which requires either fluctuations about the mean field equations
or a tail in the initial droplet size distribution, from ``spinodal" coarsening
in which a typical droplet is locally unstable.
A weaker condition for stability, previously suggested
by Kabalnov et al., is sufficient only to prevent ``spinodal'' coarsening
and is best viewed as a condition for metastability.
The coarsening of unstable emulsions is considered, and shown at
long times to resemble that of ordinary emulsions (with no trapped
species), but with a reduced value of the initial volume fraction of
dispersed phase.
We discuss the physical principles relevant to the stability of emulsions
with trapped species,
describing how these may be exploited to
restabilise partially coarsened emulsions and to ``shrink''
previously formed emulsion droplets to form ``miniemulsions''.

\end{abstract}

\renewcommand{\thefootnote}{\roman{footnote}}

\section{Introduction}

A phenomenological model for the growth of small droplets or
precipitates from a supersaturated phase, was first
investigated by Lifshitz and Slyozov\cite{1} in $1961$. They
considered a two-phase system consisting of a vanishingly small
volume fraction of droplets of a dispersed phase, evenly distributed
throughout a second, continuous phase. The model is extremely
general and forms an explanation for a number of different physical
phenomena. Among these is the coarsening of emulsion droplets made
of a liquid of low but finite solubility in the surrounding
continuous phase\cite{2,3,4,5,6}. The Lifshitz-Slyozov model is
based on a mean-field description of the evaporation/condensation
process (Ostwald ripening), which
ignores the contributions of {\em coalescence} (droplet fusion).
Depending upon the  nature of the local interactions between
droplets, their mobility, and the solubility of the dispersed phase
within the continuous one, this can be a good approximate
description of the coarsening of emulsion droplets. The
effects of coalescence are ignored in this paper.
The absence of
coalescence may require the presence of surfactant to stabilize
the emulsion droplets (for example by imparting a surface charge).
We do not treat surfactant dynamics explicitly, assuming in effect
that the surfactant transports rapidly through the continuous medium
to ensure a constant surface tension of droplets at all times.

Recent work\cite{7,8,9,Taisne} has considered how the process
differs when
there are (disregarding surfactant) three species present\cite{2,3,4}.
In
particular, Kabalnov et al.\cite{7} considered a limiting situation
where the third species is completely immiscible with the continuous
phase, and is therefore effectively {\em trapped} within droplets.
Kabalnov et al. gave quantitative criteria for the stabilisation of
an emulsion by trapped species under the condition that the initial
droplets were narrowly distributed in both size and composition.
Related ideas to these can be found in the literature on
meteorology\cite{10,11,12,13,14}, where (for example) the stability
of sea-mists and/or smog is in part attributed to the inhibition
of coarsening by the entrapment of salt or other non volatile
species.

In this paper, we consider the effects of trapping a third species
within emulsion droplets in greater detail. (Although we speak only
of emulsions, several of our ideas relating to the stabilisation of
droplets and precipitates to coarsening  may be useful in other
contexts\cite{14,15,16}.) We reexamine the condition for stability
proposed in Ref.\cite{7}, and conclude that this is, for monodisperse
droplets, the requirement for {\em metastability}, rather
than full stability.
We give a more stringent condition for full stability,
which allows for fluctuations about the mean-field growth rate of
droplets of a given size. Our stability condition is then extended
to the case of an arbitrary initial distribution of droplet sizes and
compositions. We note that this stability condition generalises
to nonideal equations of state, such as would be needed to describe
trapped salt, for example. The classification of
stable and metastable regimes is, for emulsions, somewhat different
from the one which applies to a single droplet surrounded by
vapour, as was studied (with a trapped species present)
in Ref.\cite{8}.

These arguments, which are described in Sections 2-4, are both
general and rigourous. Indeed, in the absence of coalescence
processes, the problem of the stability of dilute emulsions with
trapped species is found to be analogous to the finding of phase
equilibrium in multicomponent mixtures. The only difference is that
some extra thermodynamic variables are required, to  describe the
fixed number of droplets and their trapped contents. These variables
would not be conserved if either coalescence were to occur, or the
trapped species were able to diffuse through the continuous phase at
a finite rate. (The latter case is briefly
discussed, at the end of Section 4.) But if these processes are slow
enough, our methods
allow the state of the system, after the remaining degrees of
freedom have attained equilibrium, to be determined.

In Section 5, we consider the effect of a trapped phase on the
kinetics of coarsening, in a system which does not obey the
criterion for stability derived earlier. Approximate solutions for
the droplet size distributions are given, and the physically
motivated approximations upon which they are based are explained.
Our analysis supports the view of Ref.\cite{7}, that in the unstable
(coarsening)
growth regime, the long time dynamics is essentially the same as
given by the Lifshitz Slyozov description.

In Section 6, several implications and possible applications of
these  results are considered, including possibly novel methods for
forming stable emulsions of small droplet size, and ways of
arresting or reversing coarsening after it has already begun.
Section 7 gives a brief conclusion and outlook for further work.

\section{Kinetics of Droplet Growth} \label{growth_kinetics}

We consider a vanishingly small volume fraction of droplets and
assume   that droplets are in equilibrium with their local spatial
environments.  We also assume that the ensemble of droplets
interacts via diffusion of the dispersed-phase species through the
continuous medium, and that this diffusion is rate-limiting (it is
much slower than the rate at which the dispersed-phase species may be
absorbed into a droplet from immediately outside). Each droplet's
growth rate is then determined by the difference in concentration of
dispersed-phase species immediately outside its surface and the
(average) far-field concentration within the continuous phase. These
assumptions reduce the calculation to (i) a single body equilibrium
problem, to find the mean growth rate of a droplet in terms of its
size and the ambient supersaturation; and then (ii) a many-body
problem to determine the droplet size distribution as a function of
time. Since in stage (ii) each droplet interacts with the others
only through the ambient supersaturation, the problem is tractable.

The method and assumptions
just described are basically the same as those first used by
Lifshitz and Slyozov\cite{1}
to obtain the average
rate of droplet growth and an asymptotic solution for the droplet
size distribution. Although other methods have been used to correct
and improve those used by Lifshitz and Slyozov, the resultant
corrections are small and the resulting asymptotic solutions
essentially the same. For a clear exposition of the main ideas see
Bray\cite{17}. A review of recent theoretical and experimental work
is given by Voorhees\cite{18}, and includes an overview of
theoretical work on systems at finite droplet volume fractions.

\subsection{Local Equilibrium of a Droplet}

We treat all components of the emulsion as incompressible fluids, in
which case the Helmholtz free energy of a dispersed-phase drop is
given by
\begin{equation} F_d=F_b+{\sigma}4{\pi}R^2
\end{equation} where $F_b$ is the free energy of the same amount of
bulk liquid, $\sigma$ is the surface tension, and
$R$ is the droplets radius.  This results in a chemical potential
given by
\begin{equation} {\mu}_d= {\mu}_b +\frac{2{\sigma}v_b}{R}
\end{equation} where $v_b$ is the volume of a single molecule, and
${\mu}_b$ the chemical potential, in a bulk liquid of the dispersed
phase.

Some of the dispersed-phase species does not reside in droplets but
is instead molecularly dissolved in the continuous phase at
concentration $C$. The chemical potential of such molecules is
\begin{equation} {\mu}_c = {\mu}_{c0} + k_BT\ln C
\end{equation} where $\mu_{c0}$ is a reference value. (This assumes
that the molecular solution is dilute enough to behave ideally.) At
equilibrium ${\mu}_d={\mu}_c$ so, immediately above a droplet
surface, the concentration $C(R)$ obeys
\begin{equation} k_BT\ln (C(R))=\frac{2{\sigma}v_b}{R} + {\mu}_b -
\mu_{c0}
\end{equation} At a flat interface with radius of curvature
$R\rightarrow \infty$, we have $k_BT\ln(C(\infty))= {\mu}_b -
\mu_{c0}$, so that we may write
\begin{equation} C(R) =
C(\infty)\exp\left(\frac{\Delta\mu}{k_BT}\right)
\label{cint}
\end{equation} where $\mu_{c0}$ has been eliminated and
\begin{equation} {\Delta\mu} = {\mu}_d - {\mu}_b = {2\sigma v_b\over
R}
\label{baredeltamu}\end{equation} For small values of
${\Delta\mu}/k_BT$, as are generally maintained during the late
stages of the coarsening process, we may expand Eq.\ref{cint} to
obtain
\begin{equation} C(R) \simeq C(\infty) \left( 1 +
\frac{2{\sigma}v_b}{k_BTR} \right) \label{cinlin}
\end{equation} In fact, for typical emulsion systems near room
temperature with interfacial tensions $ \leq 10^{-1}$Nm$^{-1}$ and
initial droplets $\sim 0.1 {\mu}$m or larger in size (as would
arise for emulsions made by mechanical agitation) the
approximation
${\Delta\mu}/k_BT \ll 1$ is likely to be valid throughout the
coarsening process.

\subsection{Rate of Droplet Growth}

Under the conditions studied here $\overline{c}$, the average
concentration of dispersed-phase species dissolved in the continuous
phase, will itself be close to
$C(\infty)$. Hence  the reduced supersaturation of dispersed-phase
species, which we define as
\begin{equation}\epsilon=(\overline{c} -
C({\infty}))/C({\infty})\end{equation} will be small. This means
that droplet sizes change slowly on the time scale of relaxation of
the diffusion field: a steady-state approximation may then be made
for the concentration profile around a droplet. This entails
replacing the diffusion equation
${\partial}c/{\partial}t=D{\nabla}^2c$ with ${\nabla}^2c=0$.
Imposing as boundary conditions $c(R) = C(R)$ (the equilibrium value
at the droplet surface) and
$c(\infty) = \overline{c}$ (the mean value far away), we obtain the
steady-state profile
\begin{equation} c(r) = R \left( \frac{C(R) -
\overline{c}}{r} \right) + \overline{c}
\end{equation} for the concentration field $c(r)$ at a distance $r$
from the centre of a droplet of radius $R$.

Since we assume that the process is diffusion-limited, the rate of
droplet growth at the surface is given by the incident flux as
\begin{equation}
\frac{dR}{dt} = v_bD \left. \frac{{\partial}c}{{\partial}r}
\right|_R = v_bD\left(\frac{\overline{c} - C(R)}{R} \right)
\end{equation} Using the linearised form (\ref{cinlin}) for $C(R)$
we obtain
\begin{equation}
\frac{dR}{dt} = \frac{Dv_bC(\infty)}{R}
\left(
\epsilon  -
\frac{2{\sigma}v_b}{k_BTR}  \right)
\label{ugrowth}\end{equation} If we define reduced variables
$R'=Rk_BT/(2{\sigma}v_b)$,
$t'=tDC(\infty)k_B^2T^2/4v_b{\sigma}^2$ and and a reduced growth rate
$U(R',\epsilon)=dR'/dt'$ then Eq.\ref{ugrowth} simplifies
to\cite{17}
\begin{equation} U(R',\epsilon)= \frac{\epsilon}{R'} -
\frac{1}{R'^2}\label{growthbare}
\end{equation} This shows the balance between condensation (first
term) and the evaporation driven by the Laplace pressure effect
(second term). Note that no finite droplet size can be stable
against evaporation at long times, when the supersaturation
$\epsilon$ tends to zero. Accordingly, emulsion droplets of a single
species with finite diffusivity will always coarsen.

\section{The Effect of a Trapped Species}\label{trapped} Molecules
of a third species which are entirely immiscible with the continuous
phase, are now considered to be present within droplets. Such
molecules are effectively {\em trapped}. This has the immediate
consequence that no droplet of the dispersed phase can ever entirely
evaporate. So the trapped phase results in the additional
thermodynamic constraint that the {\em total number of droplets}
remains constant in time, with a value determined by the initial
conditions.

We assume for simplicity that the trapped species may be treated as
dilute within each droplet. (This assumption is relaxed in Section
\ref{other} below). In this case an additional term which
corresponds physically to the osmotic pressure of the trapped phase,
will contribute to the droplets chemical potential. Indeed it is
easily confirmed\cite{7} that Eq.\ref{baredeltamu} is modified to
\begin{equation} {\Delta}{\mu}=\left[\frac{2\sigma}{R}
-\frac{{\eta}k_BT}{(4/3{\pi})R^3}\right]v_b \label{deltamu}
\end{equation} where $\eta$ is the number of trapped particles in
the droplet. Thus, the osmotic pressure of the trapped species
(second term) competes directly with the Laplace pressure (first
term); the latter favours fewer, larger droplets whereas the osmotic
pressure favours a uniform droplet size (at least if all droplets
have the same
$\eta$).

\begin{figure}[htb]
\begin{center}
        \epsfysize=80mm
        \epsfxsize=100mm
        \leavevmode
        \epsffile{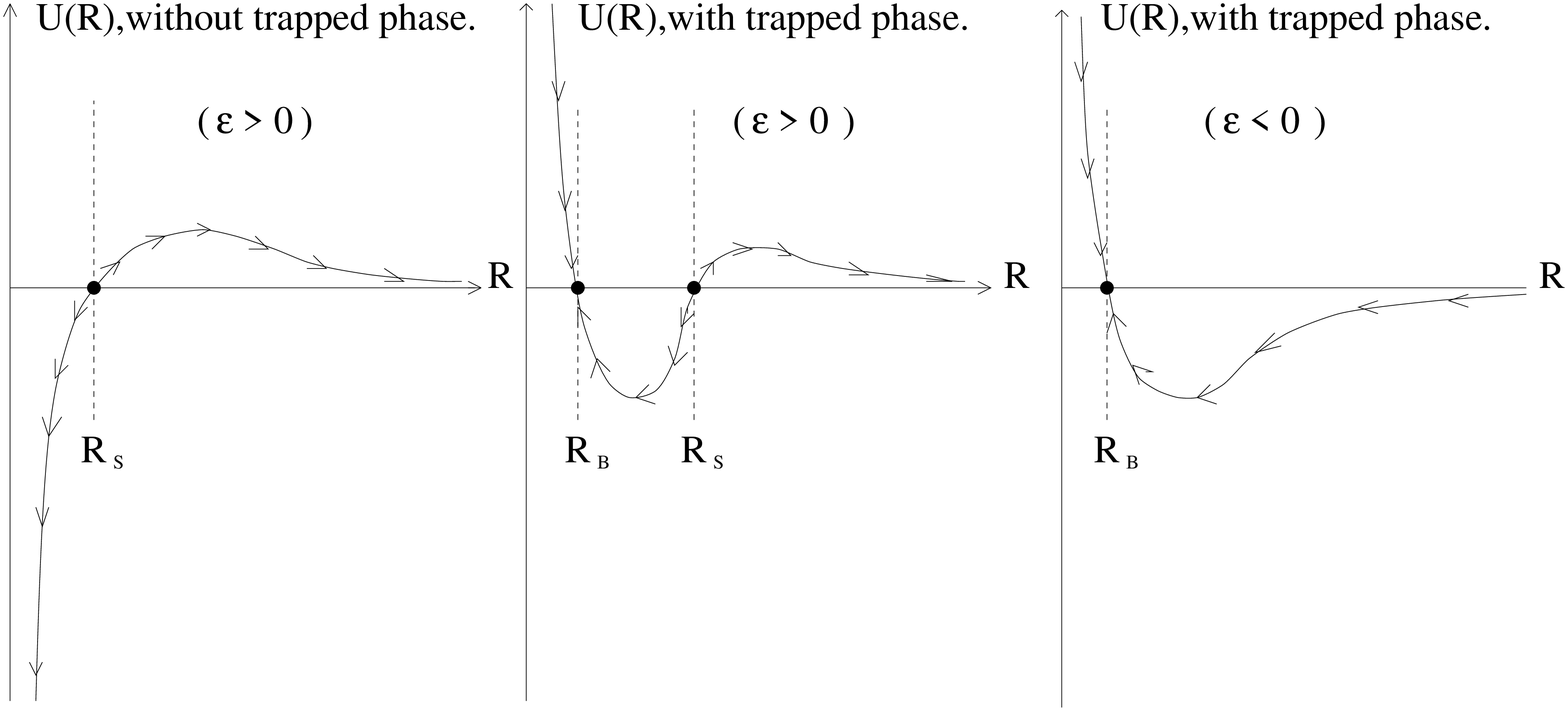}
\end{center}
\caption{A schematic diagram representing how the presence of a
trapped   phase introduces a stable droplet size. Droplets of a
given size move in the direction shown by the arrows. Without
trapped species there is one unstable fixed point at $R_S$. With a
trapped
species a new, stable, fixed point appears at $R_B$ due to the
competition
between Laplace and osmotic pressure. If there are trapped species
and $\epsilon$ is negative then there is again
only one fixed point, but it is the stable one at
$R_B$. Formulae for $R_S$, $R_B$ appear in
Section $4.2$ below.} \label{first}
\end{figure}

Note that the second term diverges as $R \rightarrow 0$; although
the trapped species cannot be treated as dilute in this limit, the
above formula is already enough to prevent a droplet from ever
evaporating completely.
Using Eq.\ref{deltamu} to determine a growth rate $dR/dt$ we find that
Eq.\ref{ugrowth} becomes
\begin{equation}
\frac{dR}{dt} = \frac{Dv_bC(\infty)}{R}
\left(
\epsilon  - \frac{2{\sigma}v_b}{k_BTR} +
\frac{{\eta}v_b}{(4{\pi}/3)R^3}
\right)
\end{equation} When $\eta=0$, a characteristic length scale
$R_{\epsilon}$ can be defined by
\begin{equation} \label{R-epsilon}
R_{\epsilon}=\frac{2{\sigma}v_b}{{\epsilon}k_BT}
\end{equation} which gives a critical size above which droplets will
grow in size and below which droplets will dissolve. When $\eta > 0$
we may define another characteristic length scale $R_B$ by that at
which the osmotic and Laplace pressures are in balance, which gives
\begin{equation} R_B = \left(3\eta k_BT\over 8\pi \sigma
\right)^{1/2}
\end{equation} Hence the growth rate may be written in the more
informative manner
\begin{equation} \label{dRdt}
\frac{dR}{dt} = \frac{Dv_b^2C(\infty)2{\sigma}}{Rk_BT}
\left(
\frac{1}{R_{\epsilon}}  - \frac{1}{R} + \frac{R_B^2}{R^3}
\right)
\end{equation} As shown in figure \ref{first}, when $\eta > 0$ a
second zero at $R \sim R_B$ is introduced into the growth rate in
addition to the original one. The stability of the
new fixed point at $R \sim R_{B}$ suggests the possibility of
forming stable emulsions; but since the curves depend on
$\epsilon$, which is time-dependent and need not even tend to zero at
long times, the ultimate behaviour of the system is unclear. This is
considered further in the following Section.

It is possible to define a growth velocity $U(R',\epsilon)$ in terms
of dimensionless variables, as
\begin{equation} U(R',\epsilon) = \frac{\epsilon}{R'} -
\frac{1}{R'^2} +
\frac{R_B'^2}{R'^4} \label{growth}
\end{equation} where $R'$,$t'$ are as defined in Section
\ref{growth_kinetics} above, and $R_B'^2=3\eta
k_B^3T^3v_b/32\pi\sigma^3$. This form of the growth rate is used in
Section \ref{coarsening} when considering the dynamics of unstable
distributions which coarsen.

\section{Formation of Stable Emulsions}\label{FSE} In discussing
stability
criteria, we assume a monodisperse initial condition with an ideal
equation of state for the trapped species within the droplets. These
assumptions will be relaxed in Sections \ref{poly} and \ref{other}
below.

\subsection{Equilibrium at Fixed Droplet Number}\label{etafix}
\begin{figure}[htb]
\begin{center}
        \epsfysize=65mm
        \epsfxsize=100mm
        \leavevmode
        \epsffile{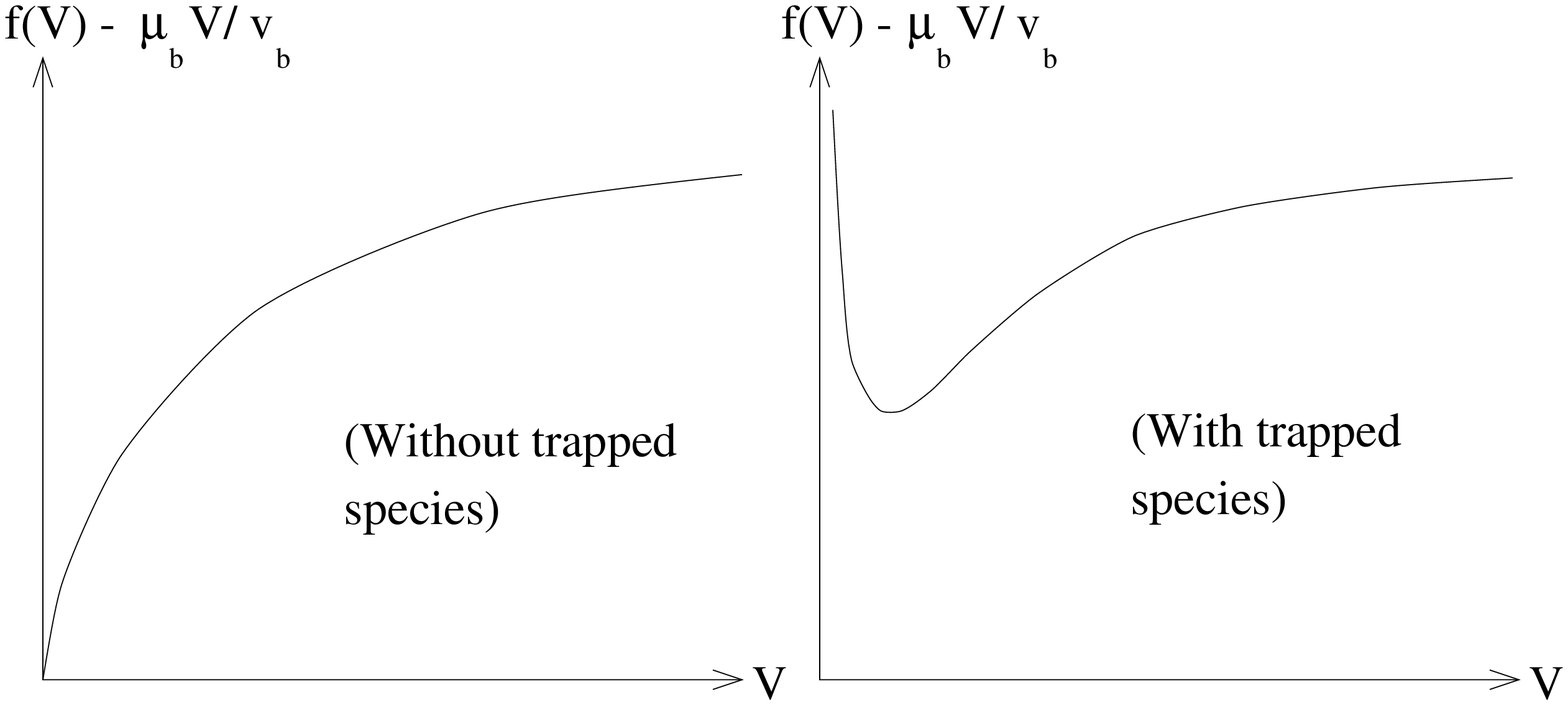}
\end{center}
\caption{Comparison of curves of $f(V)-\mu_bV/v_b$ without (left) and
with (right) trapped species.} \label{second}
\end{figure}

Consider an initially monodisperse emulsion in which the number of
trapped particles
$\eta$ is identical in each droplet. Since the emulsion droplets are
treated as macroscopic objects their entropy of mixing is
negligible, and may be ignored. The total free energy density of the
system, at some later time, may be then written
\begin{equation} F = \int dV\,n(V)f(V,\eta)
\end{equation} where $n(V)$ is the number density of droplets of
volume $V$ and $f(V,\eta)$ is the free energy of such a droplet
containing
$\eta$ trapped molecules. For an ideal trapped species,
$f(V,\eta)$ takes the form
\begin{equation} f(V,\eta) = \mu_bV/v_b + 4\pi\sigma(3V/4\pi)^{2/3}
- \eta k_BT\ln V \label{ideal}
\end{equation} where a constant has been suppressed. The curves with
and without trapped species are shown in figure \ref{second}.

The free energy density $F$ is to be minimised subject to the
constraints that $\int dV\, n(V) = n_0$ (the initial number of
droplets is held constant), and $\int dV\, V n(V) =
\phi = n_0\overline{V}$, representing a fixed total volume fraction
$\phi$ of dispersed phase\cite{phifoot}. The latter ignores a contribution from
the dispersed-phase species that is solubilized in the continuous
phase; for emulsions of low solubility, prepared by mechanical
dispersion (rather than by quenching from a homogeneous mixture at
high temperatures) the latter is always negligible. Note that the linear term,
$\mu_bV/v_b$ in $f(V)$ is irrelevant to this procedure
and, for clarity, we have subtracted it from the curves shown in
Fig.\ref{second} and similar curves appearing below.

These rules are
precisely analogous to those for constructing the equilibrium state
of a system whose free energy
$f(\lambda)$ depends on a composition variable $\lambda$. Such a
system can separate into volumes $v(\lambda)$ of phases with
different compositions $\lambda$; in that case,
$F=\sum_\lambda v(\lambda)f(\lambda)$ is minimised subject to the
constraints $\sum_\lambda v(\lambda) = v_0$ (the total volume of the
system is fixed) and $\sum_\lambda v(\lambda)\lambda = v_0\lambda_0$
(the total amount of species
$\lambda$ is conserved). Here the subscripts $0$ describe a
hypothetical homogeneous state.

In this analogy, each value of the droplet size $V$ corresponds to a
``phase" of the system, and $n(V)$ corresponds to the ``volume" of
such a phase present in the final state. A monodisperse emulsion
represents a ``single phase". The minimisation of $F$ is therefore
exactly as one would perform to find phase equilibrium in a binary
fluid with the function
$f(\lambda)$ replacing $f(V,\eta)$: the usual construction for this
is to seek common tangencies whereby $F$ can be lowered by phase
separation. (In this analogy, phase separation corresponds to the
formation of droplets of more than one size.) The volumes
$v(\lambda)$ are determined by the lever rule\cite{dehof}, and, in
principle, the same rule would apply here to calculating
$n(V)$ under conditions where more than one droplet size was present
in equilibrium. Notice that the
correspondence works only for the curve
$f(V,\eta)$; no similar construction applies to
$f(R,\eta)$, which is the form more usually considered in
the literature \cite{7,8}. Also, note that any prediction of a
``single phase" ({\em i.e.}, a monodisperse emulsion) is, in
principle, subject to a small spreading of the size distribution
arising from the entropy of mixing of the emulsion droplets
themselves (rather than of the contents within one droplet), which we
have neglected.

\subsection{Stability Criterion for Monodisperse
Emulsions}\label{etafix2} According to the above argument, if
$\eta$ is the same for all droplets (as will usually be nearly true
if the initial droplet distribution is monodisperse) the equilibrium
state
of the system at fixed droplet number can be found by inspection of
the
$f(V,\eta)$ curve. A monodisperse emulsion can be stable only if the
corresponding $V$ lies in a part of the curve of positive curvature;
if this is not the case then the free energy may be reduced (at
fixed total number of droplets and fixed $\eta$ in each drop) by the
monodisperse distribution becoming polydisperse, and coarsening will
occur. However, though necessary for stability, positive curvature
is not sufficient to prevent coarsening, since even when the
curvature is positive at a point $V=V_0$,
it may be possible to find a lower free
energy by constructing a common tangent on the $f(V)$ curve
which lies below $f(V_0)$. The shape of the $f(V)$ curve
shown in figure \ref{second} dictates that any such tangency
must connect the point at $V \rightarrow \infty$ to the absolute minimum
of $f-\mu_bV/v_b$; the latter arises at $V=V_B(\eta)$ and is discussed
further below.

The weaker of the two stability conditions (positive curvature) is
precisely that given by Kabalnov et al\cite{7}:
\begin{equation} V\le V_{S}(\eta)=\left(
\frac{3{\eta}k_BT}{2{\sigma}}
\right)^{3/2}
\sqrt{\frac{4{\pi}}{3}}
\label{partstab}
\end{equation} Hence Kabalnov et al\cite{7} reasoned that a
sufficiently monodisperse initial distribution with droplet size
$V_0<V_{S}\equiv 4\pi R_{S}^3/3$ would be stable. However, the above
argument shows this is actually a criterion for metastability. In
other words, the criterion of Ref.\cite{7} actually separates
initial distributions which coarsen immediately, from those where
coarsening requires some fluctuation to bring it about. Since every
droplet has a slightly different environment, such fluctuations are
invariably present,
though they are not included within the
theory of Lifshitz and Slyozov.
The kinetic mechanism for fluctuation-induced coarsening
is discussed in the following Section
\ref{etafix3}.  The extent of the metastable region, and the
likelihood of a suitable fluctuation actually occuring, are discussed
further in Section
\ref{meta_features}.

In any case, the above thermodynamic analogy shows that full
stability in fact arises for initially monodisperse emulsions if and
only if
\begin{equation} V_0\le V_B(\eta) =
\left(\frac{{\eta}k_BT}{2\sigma}\right)^{3/2}
\sqrt{\frac{4\pi}{3}}
\label{fullstab}
\end{equation} This corresponds to the requirement that the
initial state lies to the left of the absolute minimum in the
function
$f(V,\eta)-\mu_bV/v_b$. Everywhere to the right of this minimum, a
lower global free energy can be constructed by a common tangency
between $V_B$ and (formally)
$V=\infty$. This corresponds to a final, coarsened state in which
a monodisperse emulsion of droplet size
$V_B=4\pi R_B^3/3$ and droplet number density $n_0$ coexists with an
``infinite droplet", which can be interpreted as a bulk volume of
the dispersed-phase species.

\begin{figure}[htb]
\begin{center}
        \epsfysize=80mm
        \epsfxsize=120mm
        \leavevmode
        \epsffile{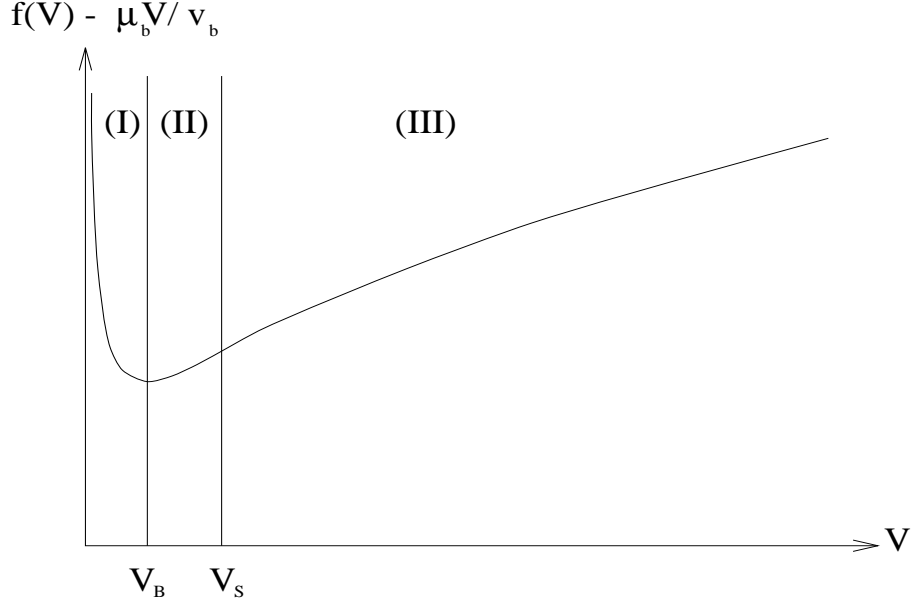}
        \end{center}
        \caption{The three different stability regimes.}\label{spincurve}
\end{figure}

The subscripts $B,S$ for $V_{B,S}$ and $R_{B,S}$ can be taken to
denote ``balance" (between osmotic and Laplace pressures in a drop)
and ``stability" (in the sense of Kabalnov et al\cite{3}). However,
in view of the thermodynamic discussion, it might be better to
interpret them as ``binodal" and ``spinodal". Indeed, depending upon
the value of $V_0$, any monodisperse distribution lies in one of
three possible regimes: see figure \ref{spincurve}. In regime I, the
emulsion is fully stable under coarsening dynamics (though not, of
course, under coalescence). In regime II it is metastable. In regime
III it is locally unstable and will coarsen immediately. The
metastable region II is both mathematically and physically analogous
to that between the binodal and the spinodal lines governing phase
coexistence in a binary fluid.

\subsection{Kinetic Interpretation} \label{etafix3} 
\begin{figure}[htb]
\begin{center}
        \epsfysize=80mm
        \epsfxsize=120mm
        \leavevmode
        \epsffile{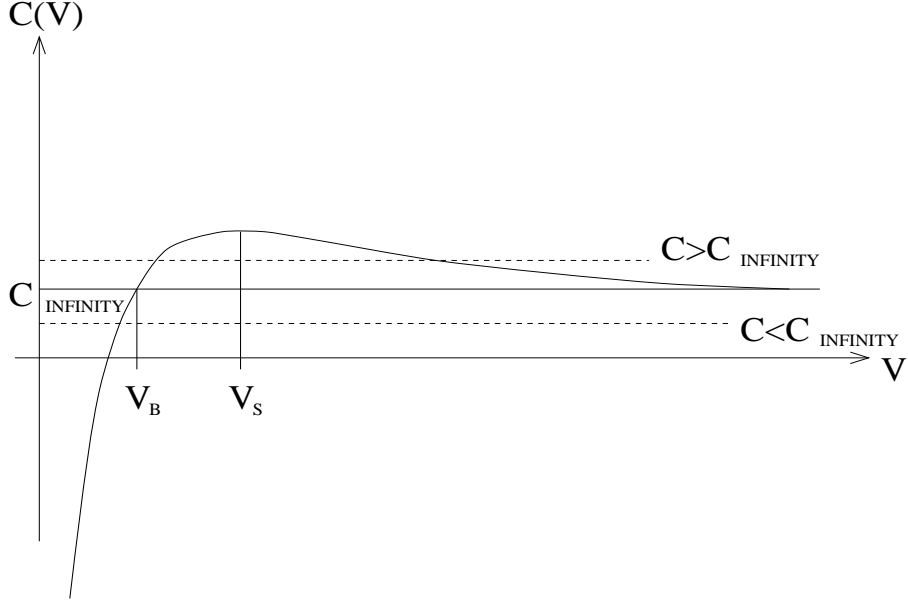}
        \end{center}
        \caption{Variation in concentration of disperse phase at a
droplets surface as a function of droplet volume.}\label{conc}
\end{figure}

The above
thermodynamic argument is quite formal, so it is useful to interpret
it in kinetic terms. Consider a monodisperse distribution which is in
equilibrium with its ``vapour" ({\em i.e.}, the dissolved fraction
of the dispersed-phase species). Let the radius of all but one
droplet be (say)
$R_0<R_S$, but suppose a single larger droplet is present, of radius
$R$. The growth of this droplet is determined by whether
$C(R,{\eta})$ is larger or smaller than
$\overline{c}$. Since the remaining droplets are in equilibrium,
$\overline{c}$ will be equal to
$C(R_0,{\eta})$. Now consider the concentration of dispersed-phase
species
$C(R,{\eta})$ at the surface of the anomalously large droplet (see
figure \ref{conc}).

Distributions in regime II have $R_B < R_0 < R_{S}$ and an average
concentration of disperse phase $\overline{c}(R_0,{\eta}) >
C({\infty},0)$. Hence if a perturbed droplet is sufficiently large,
its surface concentration will be below the ambient
level (see figure \ref{conc}) and the droplet will grow as a result of
diffusive flux through the continuous phase, at the expense of smaller
droplets.
Distributions in regime I ($R_0 < R_B$) have
$\overline{c} < C({\infty},0)$ and there is no size of droplet
larger than $R_0$ for which sustained growth is possible. Such a
droplet will instead redissolve to rejoin the equilibrium droplets
at size $R_0$.

Returning now to distributions in regime II ($R_B < R_0 <
R_{S}$), the critical size $R_C$ above which a droplet will become
unstable and start to
grow is given by the larger root of $dR/dt=0$, where $dR/dt$ is given by
Eq.\ref{dRdt}. This gives
\begin{equation} R_C=R_{\epsilon}\left( 1 -
\left(\frac{R_B}{R_{\epsilon}}\right) -
2\left(\frac{R_B}{R_{\epsilon}}\right)^2 -\dots \right)
\label{critsize}
\end{equation} where $R_{\epsilon}$ is as in Eq.\ref{R-epsilon} and
depends upon the supersaturation $\epsilon$. Since the initial
distribution of droplets is considered to be monodisperse with size
$R_0$ and in equilibrium, then $dR/dt|_{R_0}=0$ and $R_{\epsilon}$ may
be determined as
\begin{equation}
R_{\epsilon} = \left( \frac{R_0^3}{R_0^2-R_B^2} \right)
\label{destabsize}
\end{equation}
This suggests that, for a monodisperse initial state, the
nucleation time for the coarsening process to begin (requiring
nucleation of a droplet of size $R_C$) can become very large as
$R_0$ approaches
$R_B$.  On the other hand, nucleation can occur immediately if there
is any slight tail to the initial size distribution, extending
beyond $R_C$.

The distinction between nucleation due to the existence of a tail in the
initial size distribution which extends to abnormally large droplets, 
and nucleation resulting from fluctuations in growth rates, 
is similar to that between heterogeneous and homogeneous
nucleation in conventional phase equilibrium. In this analogy the tail
of the droplet size distribution (which may be negligible for
thermodynamic purposes), provides ``nucleation centres'' which allow
coarsening to begin.

\subsection{Laplace and Osmotic Pressure Balance} \label{laopb}
The absolute
stability requirement, $R_0 \le R_B$ (Eq.\ref{fullstab}), is
equivalent to requiring that
$C(R_0,{\eta}) \le C({\infty},0)$. However, we have for an ideal
trapped species
\begin{equation} C({R},\eta)=C({\infty},0)\left(
1+\frac{v_b}{k_BT}\left[\frac{2\sigma}{R} -
\frac{{\eta}k_BT}{(4{\pi}/3)R^3}
\right] \right)
\end{equation} Identifying $2{\sigma}/R$ with the Laplace pressure
${\Pi}_{L}$ and
${\eta}k_BT/(4{\pi}/3)R^3$ with an osmotic pressure
${\Pi}_{osm}$ of the trapped species, we see that the absolute stability
condition may be written simply as
\begin{equation}{\Pi}_{L} < {\Pi}_{osm}
\end{equation} or alternatively
${2\sigma}/{k_BT} < c_tR$ where $c_t = \eta/V_0$ is the
concentration of trapped species in the dispersed phase.
This means that if an emulsion is made from a dispersed phase of
fixed $c_t$,  a monodisperse emulsion will be fully stable only if
the initial droplet size is sufficiently {\em large}. (The same
applies to the metastability condition of Ref.\cite{3}.) This may be
a rather unintuitive result; indeed, at fixed number of trapped
particles $\eta$, as was used to discuss the $f(V,\eta)$ curves,
stable emulsions  arise only for {\em small} droplet sizes ($R_0 \le
R_B$). However, for a fixed composition $c_t$, the value of $\eta$
depends on $R_0$; and the important requirement is that the osmotic
pressure (which inhibits coarsening) of an initial droplet exceeds
its Laplace pressure (which drives it). At given $c_t$ this is true
only for large enough droplets.

\subsection{Stability of Polydisperse Emulsions}\label{poly}

The
criterion discussed above for full stability can be extended to the
polydisperse case. Let us first consider the case where the initial
state contains droplets with variable sizes
$V_0$, but exactly the same
$\eta$. This system will again find its equilibrium state at fixed
total number of droplets $n_0$ and fixed volume fraction $\phi$,
which now obeys
\begin{equation}
\phi = n_0\overline V
\end{equation} with $\overline V$ the mean initial droplet volume.
By again ignoring the entropy of mixing of the droplets (treating
them as macroscopic objects), we see that the thermodynamic
arguments developed in Section \ref{etafix} for equilibrium at fixed
droplet number apply without modification to this type of
polydispersity. Therefore, the ultimate behaviour is found simply by
substituting $\overline V$ for $V_0$ in our previous discussion.
Thus the criterion for full stability in this case is
$\overline{V}\le V_B$. It is also clear that in the stable regime
the droplet distribution will evolve under the
evaporation/condensation dynamics into a monodisperse one, whatever
its initial polydispersity. This contrasts with the other regimes,
where the final state will again have density $n_0$ of monodisperse
droplets, but of size $R_B$ and in coexistence with an ``infinite
droplet" containing the excess amount of dispersed phase
($n_0(\overline{V}-V_B)$) not residing in finite droplets.

The situation when $\eta$ is not the same for all droplets is more
complex. To find a stability condition for this case, we consider
first a more formal argument which reproduces the above result for a
single $\eta$. If the emulsion is unstable, we expect that at long
times the size distribution will split into a ``coarsening part" and
a ``stable  part". As equilibrium is approached, the coarsening part
has some average size
$\overline{V}_c$ which tends to infinity at long times. This
requires that $\epsilon\to 0$, which in turn means that the stable
part of the distribution is necessarily monodisperse, and of droplet
size $V \to V_B$. (This corresponds to the coexistence condition
between finite and an infinite droplet mentioned previously.)

Let
$n_0$, $n_{s}$ and
$n_c$ be respectively the number densities of all the droplets in
the system, of those in the ``stable part" and of those in the
``coarsening part" of the distribution; clearly $n_0 = n_{s}+n_{c}$. Then as
the system tends to equilibrium, the conservation of $\phi$ (which we define
to include the trapped species)
requires\begin{equation} n_0\overline{V} = n_{s}V_B +
n_c\overline{V}_c
\end{equation}
and hence
\begin{equation} n_0(\overline{V}-V_B)=n_c (\overline{V}_c - V_B)
\end{equation} So if
$V_B \geq \overline{V}$, then
$0 \geq n_c (\overline{V}_c - V_B)$; but since
$\overline{V}_c > V_B$, then $n_c=0$ and a coarsening part of the
distribution cannot exist in this case.

Now we consider the more general situation where there is
polydispersity, not only in the initial droplet size, but also in
the quantity $\eta$ of trapped species present in the initial
droplets. Let
$n(\eta)$, $n_{s}(\eta)$ and $n_c(\eta)$ be the number densities of
droplets which contain $\eta$ trapped particles in the full
distribution, and in its stable and coarsening parts respectively.
As the system tends to equilibrium the conservation of $\phi$ and
$n_0 = \int n(\eta)\,d\eta$ now requires
\begin{equation}
\int n(\eta)\overline{V}d{\eta} = \int
n_{s}(\eta)V_B(\eta)d\eta  + \int
n_c(\eta)\overline{V}_c(\eta)d\eta
\end{equation}
\begin{equation} n_{s}(\eta) + n_c(\eta) =n(\eta) \equiv n_0p(\eta)
\end{equation} where $p(\eta)$ is the probability of a given droplet
having $\eta$ trapped particles (which is time-independent). Hence
\begin{equation}
\begin{array}{ll} n_0\overline{V}=& n_0\int p(\eta)V_B(\eta)d\eta
\\&+ \int n_c(\eta)(\overline{V}_C(\eta) - V_B(\eta))d\eta
\end{array}
\end{equation}
which may be rewritten as \begin{equation}\label{soluble_phase}
n_0(\overline{V}-{\langle}V_B(\eta){\rangle}_{\eta}) =
\int n_c(\eta)(\overline{V}_c(\eta) - V_B(\eta))d\eta
\end{equation} So if
$\overline{V} \leq {\langle}V_B(\eta){\rangle}_{\eta}$, then
\begin{equation} 0 \geq \int n_c(\eta)(\overline{V}_c(\eta) -
V_B(\eta))d\eta
\end{equation} but since $\overline{V}_c(\eta) > V_B(\eta)$, then
$n_C(\eta)=0$ for all $\eta$, and no coarsening part of the
distribution can  exist.

In summary, a condition which is sufficient to ensure the full stability
of emulsions with an arbitrary initial distribution of sizes $V$ and
trapped species
$\eta$ is:
\begin{equation}
\overline{V} \leq {\langle}V_{B}(\eta){\rangle}_{\eta}
\label{rigor}\end{equation} where $\overline{V} = \phi/n_0$ and
$V_B(\eta)$ is as defined in Eq.\ref{fullstab}:
$V_B(\eta) =
\left({4\pi}/{3}\right)^{1/2}\left({\eta}k_BT/{2\sigma}\right)^{3/2}
$. The assumptions behind this result are: (i)
the trapped species has zero solubility in the
continuous phase and forms an ideal solution in
each droplet; (ii) coalescence is strictly
absent; and (iii) the solubility of the main
dispersed-phase species is nonzero (so that
diffusion can occur) but small enough that it
contributes negligibly to $\phi$. Subject to
these, it is a rigorous result.
It is easily established, using arguments that
parallel those of Section \ref{etafix}, that
condition Eq.\ref{rigor} is not only sufficient for full stability
but also necessary, in the sense that any distribution
which does not satisfy Eq.\ref{rigor} can lower its free
energy at fixed droplet number by a nucleated (if not a spinodal)
coarsening process.
As noted in
Section \ref{etafix2} for the monodisperse case,
for an emulsion which is only slightly unstable
and whose size distribution does not have a tail
extending to large droplets, the nucleation time
before coarsening begins may be very long. If
there is a tail, a long induction time is not
expected.

Notice that our rigorous condition Eq.\ref{rigor} involves
calculating $\langle V_B(\eta) \rangle_{\eta}$ by averaging over the
probability distribution of the trapped species. This quantity is
not the same as $V_B(\langle\eta\rangle)$; indeed for $\alpha\geq1$
one has the general inequality
\begin{equation}
\int_0^{\infty} \eta^{\alpha} p(\eta) d\eta \geq
\left(\int_0^{\infty}
\eta p(\eta) d\eta\right)^\alpha
\end{equation}
Since $V_B\sim\eta^{3/2}$ ($\alpha = 3/2$) we find that $\langle V_B(\eta)
\rangle_{\eta} \ge V_B(\langle
\eta\rangle)$. Hence the approximate stability requirement
$\overline{V}\le V_B(\langle \eta\rangle)$, based on the {\em mean}
trapped particle number, underestimates the maximum initial droplet
size. Accordingly this condition is sufficient, but not necessary, to ensure
full stability.

\subsection{The Metastable Regime (II)}\label{meta_features}
For a polydisperse system the condition for full stability is clear (see above),
whereas that for metastability is less
obvious. Even when all
droplets have the same $\eta$, metastability may depend on the details
of the initial droplet size distribution. Certainly, if this has no
upper limit ({\em i.e.}, there is a finite density of droplets above
any given size), then the largest droplets will serve as nuclei for
coarsening. As a qualitative rule, one can apply the argument of
Section \ref{etafix3} in a mean-field approximation, whereby an
anomalously large droplet is considered to be exchanging material
with a set of others, which for simplicity we treat as having a
single size
$\overline{V}$. This enables a critical radius $R_C$ to be estimated
from Eq.\ref{critsize} by replacing the length scale
$R_{\epsilon}$ with
\begin{equation}
R_{\epsilon} \simeq \left( \frac{3}{4{\pi}} \right)^{1/3} \left(
\frac {\overline{V_0}} {\overline{V_0}^{2/3}-\overline{V_B}^{2/3}}
\right)
\end{equation}
As before, if the initial size distribution contains
{\em any} droplets larger than $R_C$, coarsening can be expected.

If $\eta$ varies between droplets, things are still more
complicated, since nucleation is likely to involve droplets of
larger than average $\eta$, as well as larger than average size.
Although we still expect three regimes (fully stable, metastable and
unstable) corresponding to those discussed in Section \ref{etafix3}
for the monodisperse case, the boundary between the metastable
and unstable regimes may
have a complicated dependence on the initial distribution of
droplets and trapped species. This contrasts with the very simple
criterion for full stability, Eq.\ref{rigor} which applies for arbitrary
initial conditions.

Note that the dynamics we propose for nucleation in
the metastable region is peculiar to emulsions. For example, Reiss and
Koper\cite{8} discussed the equilibrium of a single drop within a
uniform environment at  fixed supersaturation. They concluded that for
their problem nucleation dynamics
were likely to be extremely slow, and deduced a criterion for stability
corresponding to that given by Kabalnov et al. \cite{7}. This does not
contradict our own conclusions, because Reiss and Koper consider a {\em 
single drop} as opposed to a population of droplets.
For an emulsion in regime II, only the largest drop present need exceed 
the nucleation threshold to initiate coarsening, and in many cases a
large enough droplet would be present in the initial droplet size
distribution. What is more, in a real emulsion with finite volume
fraction, variations in the local environments experienced  by
individual droplets will be a source of dynamical fluctuations which may 
far exceed the purely thermal fluctuations considered in Ref.\cite{8}.

In summary, the stability conditions for a single droplet, and a
population of droplets, are very different. We suspect that, even in a
nominally monodisperse emulsion, nucleation rates are generally not
negligible, and hence that the only reliable criterion for stability is
the one we have given in section \ref{poly}.

\subsection{Other Equations of State}\label{other}

The thermodynamic arguments of Sections \ref{etafix} and \ref{etafix2}
generalise readily to an arbitrary equation of state for the trapped
species. Indeed, for the case where $\eta$ is the same in all
droplets, one need only replace Eq.\ref{ideal} with,
\begin{equation} f(V,\eta) =\mu_bV/v_b + 4\pi\sigma(3V/4\pi)^{2/3} +
f_t(V,\eta) \label{eosgen}
\end{equation} where $f_t(V,\eta)$ is the free energy of $\eta$
trapped particles in volume $V$ and could include arbitrary
interactions between these.
If all droplets have the same $\eta$,
the condition for stability remains that $\overline{V}\le V_B$ where
$V_B(\eta)$ is the droplet size corresponding to the absolute minimum
of $f(V,\eta)-\mu_b V/v_b$. For such droplets, the osmotic pressure
$(-\partial f_t/\partial V)_\eta$ is again in exact balance with the
Laplace pressure. The metastability criterion is again that
$f(V,\eta)$ has positive curvature. For the case where $\eta$ is not
the same for all droplets, the full stability criterion is again
Eq.\ref{rigor}.

A complication arises if the interactions between the trapped species
are attractive.  Reiss and Koper\cite{8} pointed out that
in this case a third, unstable fixed point can exist at a size smaller
than
$V_B$. They also correctly
point out that such an attraction is likely to cause phase separation
within
droplets. The latter occurs whenever $f_t(V,\eta)$ has negative
curvature (with
respect to $V$ at fixed $\eta$); the
form of $f_t(V,\eta)$ in Eq.
\ref{eosgen} must then be modified to reflect the internal phase separation, and
any negative curvature regions will then be replaced by zero-curvature ones
(corresponding to tie lines).  In this case, because of the surface tension
contribution in Eq.\ref{eosgen}, there can arise an additional minimum in
$f(V,\eta) -\mu_bV/v_b$ which could lead, for example, to stable bidisperse
emulsions for some range of initial conditions. The same can arise without
intra-droplet phase separation,  if $f_t(V,\eta)$ has a small enough
positive curvature for this to be outweighed by the
negative contribution from the surface tension term in Eq.\ref{eosgen}. We
leave a detailed discussion of these cases for future work.

These results are sufficient, for example, to deal with the
case of a distribution containing trapped salt (where $f_t(V,\eta)$ can be
approximated by, say, the Debye Hueckel equation of state
\cite{mcquarry}) which is of interest in meteorological as well as
emulsion stability contexts \cite{10,11,12,13,14}.

In the case of repulsive interactions, at least, our  stability condition
Eq.\ref{rigor}  is also easily generalised to a situation in which there is
more than one trapped species. Letting
${\eta}_i$ be the number of particles of the ith species trapped
within a droplet, then the condition for stability generalises to
\begin{equation}
\overline{V} \leq \langle V_B({\eta}_1,{\eta}_2,\dots)
\rangle_{{\eta}_1,{\eta}_2,\dots}\label{multispec}
\end{equation} where $V_B({\eta}_1,{\eta}_2,\dots)$ is the droplet
size corresponding to the absolute minimum of
$f(V,{\eta}_1,{\eta}_2,\dots) - \mu_b V/v_b$. A proof of this
generalized stability condition follows
that given in Section \ref{poly}, with the single variable $\eta$
replaced by the list of variables $\eta_1,\eta_2,\dots$.

So far, we have not considered explicitly the role of surfactant (which is
usually present to prevent droplet coalescence in emulsions), tacitly assuming
that this merely alters the constant value of $\sigma$, the surface tension.
For a surfactant that is insoluble in the continuous phase, this surface
tension
will itself be a function of droplet size. We do not treat this case
further, but note that it could be included in Eq.\ref{eosgen} by
replacing the term in
$\sigma$ with a suitable ``surface equation of state" for the trapped
surfactant. The same applies to bending energy terms which could be
significant
for extremely small droplets.

\subsection{Behaviour when the ``trapped" phase is
partially soluble}\label{partisol}

Throughout the above we have treated the trapped species as entirely
confined in the emulsion droplets. In this Section we briefly consider
what happens if this third species is very slightly soluble. (For simplicity we
treat $\eta$ as the same for all droplets.)
Our previous classification into regimes I,II and III, though no
longer strictly applicable, remains a guide to the resulting behaviour. The
discussion that follows is related to that of Kabalnov et al.\cite{7}.

Recall that for entirely trapped species, the unstable regime (III) is
characterised by the
evolution of a bimodal distribution of droplet sizes, in which the larger
droplets coarsen by the Lifshitz--Slyozov mechanism, while the smaller
droplets adopt a size in equilibrium with the larger drops which
approaches $V_B$ as coarsening proceeds. (This scenario is examined in
more detail in Section \ref{coarsening} below.)
If the trapped phase is now made slightly soluble, the larger
drops will, as before, coarsen at a rate determined by the transport of the
majority (more soluble) dispersed phase. However, the small droplets that
remain cannot now approach a limiting size, but will themselves evaporate
at a
much slower rate governed by the transport of the ``trapped" species.
Therefore, in the unstable regime, a two-stage coarsening is expected.
In the fully stable regime (I), on the other hand, the first of these processes
(rapid coarsening of the larger droplets) is switched off. The droplet size
distribution has a single peak, which in the fully insoluble case will
approach a delta function at the initial mean size.
However with slight solubility coarsening will occur via the
Lifshitz--Slyozov mechanism, but only at a slower rate controlled by the
transport of the less soluble species.
The behaviour in the metastable regime (II) is complex\cite{7}, possibly
characterised by a ``crossover'' from coarsening controlled by the
less soluble component to coarsening controlled by the more soluble
component, and we do not pursue it here.

For a slightly soluble trapped species, the best prospect for stability
is always to avoid the {\em rapid} coarsening process associated with
the transport of
the more soluble dispersed phase component. Since there is always {\em some}
part of the size distribution that coarsens at the slower rate set by the
trapped species, the requirement is satisfied so long as the size distribution
remains single-peaked at all times. In Appendix
\ref{solubap} we show that a sufficient condition for such behaviour is:
\begin{equation}
\overline{V}(0) \leq V_B(\overline{\eta}(0)) \label{solres}
\end{equation}
where $\overline{\eta}(t)$ is the average number of trapped species in a
droplet at time $t$. Since
$V_B(\overline{\eta}(0)) \leq \overline{V_B}({\eta}(0))$, this
condition is less easily satisfied than our absolute stability condition,
Eq.\ref{rigor}, which applies for the case of entirely trapped species.
Note, however, that although this condition is sufficient, we have
been unable to find a necessary condition. In other words, it is possible that
some systems will not satisfy Eq.\ref{solres} but will nonetheless coarsen
only slowly.
The derivation\cite{phifoot} of Eq.\ref{solres}
requires $V_B({\eta})$ to increase faster than linearly
with $\eta$; this is valid for ideal mixtures and most other systems
(perhaps excluding any in which
interactions between the trapped species are attractive; see
Section
\ref{other}).

\section{Coarsening Dynamics When a Trapped Species is
Present}\label{coarsening}

\subsection{The Physics of Coarsening} \label{coarsephys}

We now consider conditions in which the requirement for stability,
Eq.\ref{fullstab}, is not satisfied. For simplicity we take the case
where all droplets have the same number of trapped particles $\eta$.
We also assume that any nucleation event required has taken place,
and study the resulting coarsening process at late times. This
entails small supersaturation $\epsilon$.
Throughout this Section we work in
the scaled variables $R'$,
$R_B'$ and $t'$ defined in sections \ref{growth_kinetics} and
\ref{trapped}, although the primes are omitted for convenience.

In the unstable case, as mentioned in
Section \ref{trapped}, the equation for static equilibrium (zero
growth) of a given droplet,
$U(R,{\epsilon}) =0$, has both a stable fixed point close to the
balanced droplet size ($R
\sim R_{B}$) and an unstable one at $R \sim 1/{\epsilon}$. The
latter will lead to coarsening.
The following arguments suggest
that the presence of the additional,
stable fixed point influences the coarsening dynamics only in a
rather simple way. At long times, the main role of a population of
stable droplets ($R\sim R_{B}$) is to effectively exclude a finite
proportion of the dispersed phase from the coarsening process. The
remaining part, whose volume fraction $\phi$ is effectively reduced,
then coarsens almost normally.

To see this, we first expand
$U(R,\epsilon)$ for small $R-R_{B}$ and small $\epsilon$ to obtain
(in reduced units)
\begin{equation} U(R)= {\epsilon}/R_{B} -
(R-R_{B})\left(\frac{2 +{\epsilon}R_B}{R_{B}^3}\right)\label{upert}
\end{equation} with a root $U=0$ at
\begin{equation} R_T(t) = R_{B} + \epsilon(t)
\left( \frac{R_{B}^2}{2} \right) + O({\epsilon}^2) \label{asymp}
\end{equation} At late times $\epsilon$ will be small and therefore,
according to Eq.\ref{upert}, the size of any droplet in the
neighbourhood of $R_B$ relaxes exponentially toward $R_T(t)$ with a
fixed decay rate $2/R_B^3 + O(\epsilon)$. At long times, this
process will be rapid compared to any coarsening of droplets of size
$R\gg R_B$; hence any non-coarsening part of the droplet
distribution comprises an effectively monodisperse population of radius
$R_T(t)$ obeying Eq.\ref{asymp}. Put differently, $R_T(t)$ is {\em
determined} by $\epsilon$, because the non-coarsening droplets are
in local equilibrium with the ambient supersaturation at all times.

Now consider drops with size $R \sim 1/{\epsilon}$, in the
neighbourhood of the second zero of $U(R,\epsilon)$ (this is the unstable
fixed point of the growth equation at given $\epsilon$). Expanding
$U(R,\epsilon)$ in $R$ about
$R=1/\epsilon$ we find
\begin{equation} U(R)= (R-1/{\epsilon}){\epsilon}^3 +
R_{B}^2{\epsilon}^4 \label{rl}
\end{equation} with a root
\begin{equation} R_L(t) = 1/\epsilon - \epsilon R_B^2 \label{rootl}
\end{equation}
 Accordingly, drops with $R(t)>R_L\sim 1/\epsilon$ will grow,
lowering the supersaturation $\epsilon$ in the system. As this
proceeds, the monodisperse small droplets ($R=R_T(t)$) will shrink
slightly to remain in equilibrium with the current supersaturation.
If one neglects this last effect, these droplets can play no role,
at late times, other than to remove from the coarsening process an
amount of material $n_0V_B$ corresponding to that required to
produce a stable monodisperse trapped emulsion in equilibrium with
an infinite droplet. The excess material
$n_0(V_0-V_B)$ then becomes concentrated in fewer, larger droplets
as coarsening proceeds by (essentially) the usual Lifshitz-Slyozov mechanism.
This scenario is confirmed in the following Section, where we obtain an
asymptotic solution for the coarsening behaviour.

\subsection{Asymptotic Analysis}\label{asymptan}

Following Lifshitz and Slyozov \cite{1} we assume a continuous
distribution of droplet sizes $n(R,t)$, with $n(R,t)\,dR$
representing the number density of droplets of radius $(R,R+dR)$, at time
$t$. Since droplets of finite size cannot suddenly
appear or disappear, the conservation of flux through droplet size
space requires \cite{1,17}
\begin{equation}\label{fluxconservation}
\frac{{\partial}n(R,t)}{{\partial}t} = -
\frac{{\partial}(n(R,t)U(R,\epsilon))}{{\partial}R} \label{coars}
\end{equation} Conservation of the volume of dispersed phase species
requires that
\begin{equation}
\frac{1}{C_{\infty}v_b} \int_0^{\infty}\frac{4\pi}{3}R^3 n(R,t)dR +
\epsilon = \epsilon_0 \label{volcons}
\end{equation} where $\epsilon=(\overline{C} -
C_{\infty})/C_{\infty}$is the degree of supersaturation and
$\epsilon_0$ its initial value.

\begin{figure}[htb]
\begin{center}
        \epsfysize=80mm
        \epsfxsize=120mm
        \leavevmode
        \epsffile{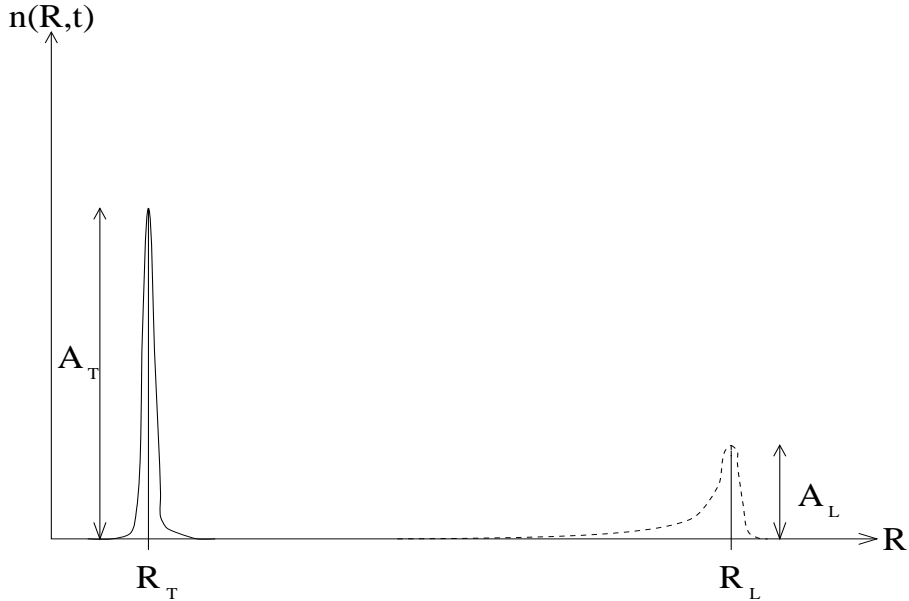}
        \end{center}
        \caption{Approximating $n(R,t)$ by functions centred on
$R_T$ and $R_L$.}\label{peaks}
\end{figure}

Taking $R_T(t)$ and $R_L(t)$ as the two roots of $U(R,\epsilon) = 0$, as
defined above, we now seek asymptotic solutions
of the form
\begin{equation} n(R,t) = A_T(R_T) f_T(R/R_T) + A_L(R_L) f_L(R/R_L)
\end{equation} where $f_T$ and $f_L$ are two different scaling
functions. That is, we assume (see figure \ref{peaks}) that at late
times the size distribution splits into two populations, each of
fixed shape in terms of an appropriate reduced size variable, and
each with some time--dependent amplitude $A$.

These two distributions are now assumed to {\em separately}  obey
Eq.\ref{coars} for conservation of flux through droplet size space.
This separation is valid so long as there is no significant range of
sizes for which both populations overlap -- which is increasingly
true at late times. However, the two populations do interact via the
supersaturation
$\epsilon$. We first
solve $U(R_L,{\epsilon})=0$ for
${\epsilon}(R_L(t))$
to obtain
\begin{equation}
\epsilon = \frac {1}{R_L(t)} \left( 1 -
\frac{R_{B}^2}{R_L(t)^2} \right)\label{invert}
\end{equation} and note that, because of this equation, terms in
$\epsilon$ and terms in $1/R_L$ are of the same order. In what
follows we take the long time limit and therefore write $\epsilon =
1/R_L + O(1/R_L^3)$. We then impose conservation of volume fraction,
and of the total number of droplets ($n_0$), to obtain $A_T$ and
$A_L$. This is done explicitly in Appendix \ref{aandb} and yields
\begin{equation} A_T =  \frac{n_0}{R_T(\epsilon)} +
O(\epsilon^3)\label{at}
\end{equation} and
\begin{equation} A_L =
\frac {\epsilon_0-n_0(4{\pi}/3)R_{B}^3/(C({\infty})v_b)}
{(4{\pi}/3)B_{0L}R_L(t)^4}  - O({\epsilon^5}) \label{ampl}
\end{equation}
where $B_{0L}$ is a moment of the scaling function $f_L$ defined
in Appendix \ref{aandb}.

Eq.\ref{fluxconservation} is
then expanded in powers of $R_L$ retaining only the lowest order
terms in $1/R_L$ (or equivalently, the lowest order terms in powers
of $\epsilon$). As shown in Appendix \ref{nl}, this results in
$R_L(t)=(3{\gamma}t)^{1/3}$ and
\begin{equation}
\begin{array}{ll}
\left[ 4f_L + Z_L\frac{{\partial}(f_L(Z_L))}{{\partial}Z_L} \right]
=&
\gamma \left[ f_L\left(-\frac{1}{Z_L^2} +\frac{2}{Z_L^3}
\right) \right.
\\ & \left. + \frac{{\partial}(f_L(Z_L))}{{\partial}Z_L}
\left(\frac{1}{Z_L} -\frac{1}{Z_L^2}
\right) \right] + O(\epsilon)  \label{brayeq}
\end{array}
\end{equation} where $Z_L=R/R_L(t)$. These equations are identical
to those solved by Bray\cite{17}, resulting in the consistency
requirement
$\gamma=4/27$, and the solution
\begin{equation} f_L(Z_L) =
\left[
\begin{array}{cc}
\frac{AZ_L^2\exp\left(\frac{-3}{3-2Z_L}\right)}
{(3+Z_L)^{7/3}(3/2-Z_L)^{11/3}} & 0 \leq Z_L < 1.5 \\ 0 & 1.5 \leq
Z_L
\end{array}
\right.
\end{equation} where $A$ is a constant which is determined by the
condition
${\int}f_L(Z_L)dZ_L=1$.

The above argument shows that the equation of motion for $R_L$ is
exactly the same as in a standard coarsening problem (with no
trapped species)\cite{17}, to the leading order in small $\epsilon$.
Accordingly the solution for $R_L(t)$, which involves seeking a
specific $\gamma$ for which the scaling distribution remains
self-consistent in the long-time limit, is also the same. The only
difference is in the {\em amplitude} $A_L$ which, as shown in
Eq.\ref{ampl}, has, to the leading order in $\epsilon$ been shifted
by a constant amount corresponding to a reduction in $\epsilon_0$.
In other words, in the long time limit the large droplets behave
{\em precisely}\cite{footsurprise} as they would for an emulsion with no
trapped
species but a reduced initial volume fraction $n_0(V_0-V_B)$.
(Obviously, the latter is assumed positive; otherwise the emulsion is
stable and will not coarsen.)

\section{Possible Applications}

\subsection{Optimal Sizing of Emulsions}
The use of ``trapped'', or less soluble species to stabilize emulsions
is widespread  in industry
\cite{19}. Such emulsions can be prepared mechanically with various average
droplet sizes. Let us assume that any trapped species is dissolved at uniform
concentration $c_t$ through the dispersed phase material. As explained
in Section \ref{laopb}, to form a stable emulsion one must ensure that
the initial droplet size is sufficiently {\em large}. Roughly speaking, this
ensures that the typical Laplace pressure $2\sigma/R$ is smaller than the
osmotic pressure $c_tk_BT$. (The quantitative version of this,
applicable to any initial size distribution, is Eq.\ref{rigor}.)
Clearly, it is important to resist the temptation to make the initial emulsion
too fine (which in ordinary emulsions might be expected to delay coarsening for
the maximum possible time). In practice, the need to avoid sedimentation (and
perhaps coalescence, which we have neglected) may set an optimal initial size
close to, but above, the absolute stability threshold,
Eq.\ref{rigor}. For the reasons discussed in Section
\ref{FSE}, the weaker ``spinodal" condition\cite{7} Eq.\ref{partstab},
even for a nominally monodisperse initial state, cannot guarantee
stability.

\subsection{Reversing the Coarsening Process}\label{reverseCor}
Consider a situation where an emulsion is prepared in an unstable
state (for example by mechanical agitation) and then starts to coarsen.
We now
ask what will happen if, after some time interval, a large number
of small droplets (containing a trapped species) are added to the
emulsion, causing the stability condition $\overline{V}
\leq {\langle}V_0(\eta){\rangle}_{\eta}$ to become satisfied for the
system as
a whole. According to the arguments of Section \ref{poly}, the emulsion is
now unconditionally stable. Therefore it will not coarsen further. What is
more, for a given total
volume fraction $\phi$ and a given population of trapped species,
the final state of the system is the unique one in which droplets of all
$\eta$ have a common chemical potential for the (mobile) disperse--phase
species. Thus the addition of the small droplets will not only prevent further
coarsening, but will in general cause previously coarsened droplets to {\em
redissolve}. Indeed, if there are no trapped species in the initial
unstable droplets, these will evaporate completely, and their material will be
entirely absorbed by the added droplets\cite{Recond}. A formal proof of
these remarks, following the lines of Section
\ref{poly}, is left to the reader.

In principle, the coarsening process can
be reversed even when it is complete. For example, a system of small
oil-in-water emulsion droplets with trapped species present can,
by the evaporation-condensation mechanism, take up oil from an excess bulk
phase of pure oil. If the stability condition  $\overline{V}
\leq {\langle}V_0(\eta){\rangle}_{\eta}$ is met by the system as a
whole\cite{footnote1},
the bulk phase of oil will disappear entirely (though obviously this process may
be slow in practice). Like the rest of our conclusions, this one applies
only if both coalescence, and diffusion of the trapped species through the
continuous phase, are strictly negligible. With these assumptions, we may
minimize the free energy with the constraint of fixed trapped species in each
drop; the above result for the equilibrium state follows immediately.

\subsection{Formation of Mini-emulsions by Shrinking}\label{miniemulsions}

``Mini-emulsions", comprised of droplets with radii of between $50$ and
$150$ nm, have many potential uses in industrial and
pharmaceutical applications\cite{19}. However their formation by
traditional mechanical methods, where droplets are formed by strongly
shearing the ingredients, is limited by the high energy required by
the process\cite{20}, and the difficulty of getting a uniform droplet size. An
alternative route is to create an emulsion of relatively large drops (whose
size is also more controllable) and then ``shrink" them to the required,
smaller size. The following describes a non-mechanical method which would
allow this to be done.

\begin{figure}[htb]
\begin{center}
        \epsfysize=50mm
        \epsfxsize=75mm
        \leavevmode
        \epsffile{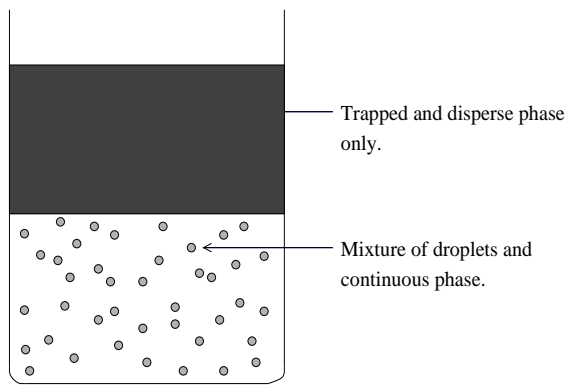}
        \end{center}
        \caption{Droplets shrink to
attain a stable size, which is in equilibrium with an excess bulk phase
that also contains a  trapped species.}\label{drying}
\end{figure}

Consider a situation in which a bulk reservoir of the dispersed-phase
species (say, oil), containing trapped species at concentration $c_b$, is
placed in  contact with a stable emulsion of oil droplets (in water, say)
each containing $\eta$ molecules of the trapped species  (see
figure \ref{drying}). This system is then allowed to reach equilibrium under the
evaporation-condensation mechanism. (To speed the process, some gentle
agitation of the emulsion might be desirable.)

A final state of equilibrium will be reached when the chemical potential of
the dispersed phase species in the emulsion equates to that in the bulk. This
requires
\begin{equation}
\Pi_{osm}^{bulk} = \Pi_{osm}^{drop} - \Pi_L
\end{equation}
where $\Pi_{osm}^{bulk,drop}$ are the osmotic pressures of trapped species in
the bulk and droplets and $\Pi_L$ is the Laplace pressure in the droplets.
For ideal solutions this condition reads
\begin{equation}
c_b = {\eta\over (4\pi/3)R^3} - {2\sigma \over R k_BT}
\end{equation}
which is an equation for the final droplet size $R$.
By increasing $c_b$ the final droplet size can be made as small as one
wishes: effectively the oil can be ``squeezed out" of the emulsion
by the osmotic action of the trapped species, now present in the bulk
oil phase at higher concentration. (This contrasts with the last
example in Section \ref{reverseCor}, where a bulk phase, containing no
trapped species, could be entirely absorbed by the emulsion droplets.)
Once the desired size of mini-emulsion is
reached, it can be removed and will remain stable.

Note that, because the trapped species is
insoluble in the continuous phase (water in this example), there is no
need for a semipermeable membrane
to prevent its transfer between the bulk oil phase and the emulsion
droplets.
Therefore, in practice, a more rapid exchange equilibrium might be reached if
the ``bulk" oil phase instead took the form of large macroemulsion droplets
which can later be separated out easily by sedimentation. Moreover, there is no
need for the trapped species in the bulk oil phase, and the trapped species in
the mini-emulsion droplets, to be identical; so long as both are insoluble in
water, the same condition for osmotic equilibrium will apply. Therefore one
can ``shrink" an emulsion containing an expensive trapped species (such as
a fragrance or drug) by contacting it with a cheap polymer solution.

The shrinking process should not only achieve small droplet sizes, but may also
allow one to reach concentrations of the trapped phase that would be
unattainable by normal means. For example, by making an emulsion of dilute
polymer solution and then shrinking the droplets, it may be possible to
achieve within each droplet a highly concentrated polymer solution, too
viscous to be dispersed mechanically in its own right.

These ideas may be relevant to various encapsulation technologies. Obviously,
the designation of ``oil" and ``water" in the above is arbitrary and
these could be any two phases.
We have assumed throughout that coalescence
is negligible, which is
commonly the case for oil-in-water emulsions so long as they contain a
surfactant (typically ionic), to give a
surface repulsion between droplets. The presence of the surfactant should not
alter our arguments, so long as it is soluble enough in the continuous phase
that the surface tension $\sigma$ does not vary between droplets. In principle,
for small enough mini-emulsions, the surfactant could also give rise to
significant bending energy terms in the free energy of a droplet, which could be
included if required (see Section \ref{other}).

\section{Conclusions}

It has long been known that emulsions containing a
sufficiently high concentration of a trapped phase will
not coarsen via the Lifshitz--Slyozov\cite{1} mechanism (Ostwald
Ripening). For practical purposes it is clearly important to know
what is ``sufficiently high" in this context; this information
can allow stable emulsions to be designed, rather than formulated by
trial and error.

In this work we have provided a general condition (Eq.\ref{rigor}) which
(in the absence of droplet coalescence) will guarantee stability against
coarsening, for emulsions of arbitrary polydispersity in both initial
droplet size and composition.
The remarkable simplicity of our
result stems from a correspondence between the thermodynamics of emulsions
at fixed droplet number and the equilibrium 
among multiphase fluids (Section \ref{etafix}).

Although presented for an ideal solution
of trapped species, our condition may be generalised to an arbitrary equation
of state (see
Section \ref{other}). In fact though, for typical parameters ($\sigma
\simeq 10^{-1}$N m$^{-1}$, $R_0\simeq 1\mu$m) stability can be achieved
with trapped concentrations of order $10^{-2}$ M so that departures from
ideality need not arise.  An analogous
condition (Eq.
\ref{solres}) was also found that ensures a relatively
long-lived emulsion even when the ``trapped"
phase is slightly soluble.

In deriving Eq.\ref{rigor} (and its simplified form, Eq.\ref{fullstab}
for the monodisperse case) we took care to distinguish ``nucleated"
from ``spinodal" coarsening.
Eqn.\ref{rigor} rigorously identifies the boundary between fully stable
emulsions
of arbitrary initial polydispersity (regime I) and emulsions that can lower
their free energy by coarsening at fixed droplet number. 
The latter can be subdivided into two
classes (regimes II and III). In regime II, coarsening can occur by
nucleation. The nucleation can arise either because of variations
in the local environments of droplets, or from the presence in the initial state
of even a {\em single} pre-existing large droplet. We believe that the latter,
in particular, can rarely be ruled out.

In contrast, spinodal coarsening (regime III)
requires neither mechanism since typical droplets are locally unstable.
We showed that for monodisperse emulsions, Eq.\ref{partstab} as proposed
by Kabalnov et al.\cite{7} actually identifies the onset of spinodal
coarsening.
If in general nucleated processes cannot be ruled out, 
condition Eq.\ref{partstab} is insufficient for stability.

The effect of trapped species on the dynamics of unstable, coarsening
emulsions was also considered. It was found that coarsening would
proceed precisely like that in an emulsion with no trapped
species\cite{1}, but with a reduced initial volume fraction
$n_0(V_0-V_B)$.

We have emphasised that the emulsions evolution is driven by the
competition between the osmotic pressure of the trapped species and
the Laplace pressure of droplets. These principles are exploited in Section
\ref{miniemulsions}, where we describe non--mechanical methods which
might allow the restabilization of a partially coarsened emulsion and the
formation of stable emulsions consisting of droplets smaller
than those attainable by traditional mechanical methods.

There is scope for further work in several directions.
For example, it would also be interesting to consider further the
detailed kinetics of emulsions
in the metastable regime (II); these can be complex, especially if the
``trapped" species is itself slightly soluble (Section \ref{partisol}).
For practical purposes in designing emulsions, however, this
regime is best avoided as we have emphasized above.  Of greater
technological importance is the case of emulsions which consist
of multiple components of varying solubility. One would like criteria for
the average
droplet composition that will maximise such an emulsion's lifetime.
We hope to return to this issue in future studies.

\vspace{0.5cm}

{\bf Acknowledgements:}  We are grateful to Alex Lips, Marcel Penders, and
especially Wilson Poon, for discussions that initiated this work. AJW
acknowledges the support of the EPSRC (UK) in the form of a Research
Studentship.

\appendix

\section{Slow coarsening condition for a slightly soluble ``trapped''
species} \label{solubap}
In this Appendix we establish Eq.\ref{solres}, which is the condition
that the size distribution remains single-peaked at all times, in a
system where the trapped particles are slightly soluble. 

We do this by considering the contrary case of a
distribution which consists of both a rapidly coarsening and
a quasi-stable part, with the rapid coarsening of the larger drops occuring at a
rate determined by the transport of the most soluble component. As
$t \rightarrow \infty$, the size $V_L$ of the larger droplets becomes
large, the supersaturation of
the disperse phases will tend to zero, and the smaller drops will
tend to  sizes $V_B(\eta)$. Here
$\eta$ may now vary among droplets
and with time.

The total number of droplets is no longer conserved but becomes
time-dependent, and Eq.\ref{soluble_phase} is replaced by
\begin{equation} n(0)\overline{V}(0) - n(t)\overline{V_B}(t) = \int
n_L(\eta,t) ( \overline{V_L}(t) - V_B(\eta) ) d\eta
\end{equation}
where $n_L(\eta,t)$ is the number of droplets at
the larger size
$V_L(t)$, and $\langle V_B(t) \rangle_{\eta}$ has been
rewritten as
$\overline{V_B}(t)$ for convenience. So if we can ensure that
\begin{equation}
\overline{V}(0) - \frac{n(t)\overline{V_B}(t)}{n(0)} \leq 0
\end{equation} then $n_L({\eta},t)=0$ and the
distribution must be single-peaked.

Conservation of the less soluble components requires that
\begin{equation} \label{cons2c} n(0)\overline{\eta}(0) =
n(t)\overline{\eta}(t)
\end{equation}
or
\begin{equation}
\frac{n(t)\overline{V_B}(t)}{n(0)} =
\frac{\overline{\eta}(0)\overline{V_B}(t)}{\overline{\eta}(t)}
\end{equation}

We next note that $V_B(0)=0$ and assume that $V_B({\eta})$ is convex, as
is the case for ideal phases where $V_B \sim {\eta}^{3/2}$.
(This ensures that if $y_2 > y_1$, then
$y_1V_B(y_2) > y_2V_B(y_1)$.) If so
\begin{equation}
\overline{V_B}(t) \geq V_B(\overline{\eta}(t))
\end{equation}
And
\begin{equation}
\overline{\eta}(0)\overline{V_B}(t) \ge \overline{\eta}(0)
V_B(\overline{\eta}(t))
> \overline{\eta}(t) V_B (\overline{\eta}(0))
\end{equation}
To obtain the second inequality, we have again used the fact that $V_B(\eta)$
increases faster than linearly with $\eta$ and also exploited the fact
that $\overline{\eta}(t)$ increases with time. (The latter follows from the
fact that the number of droplets present must decrease as time proceeds.)
Hence
$\overline{\eta}(0)\overline{V_B}(t)/\overline{\eta}(t) >
V_B(\overline{\eta}(0))$ and
\begin{equation}
\overline{V}(0) - \frac{n(t)\overline{V_B}(t)}{n(0)} =
\overline{V}(0) - \frac{\overline{\eta}(0)\overline{V_B}(t)}
{\overline{\eta}(t)} <
\overline{V}(0) - V_B(\overline{\eta}(0))
\end{equation} So if $\overline{V}(0) \leq V_B(\overline{\eta}(0))$
then
$\overline{V}(0) - n(t)\overline{V_B}(t)/n(0) < 0$ and
$n_L({\eta},t)=0$. Hence the distribution may only
contain a single peak, and will coarsen at rate determined by the less
soluble  component.

\section{Determination of $A_T$ and $A_L$}\label{aandb}
In this Appendix, we obtain results for the amplitudes $A_T$ and $A_L$
as defined in Section \ref{asymptan}. These
are
determined by the constraints of conservation of volume fraction and
conservation of the total number of droplets. We write $n(R,t) = n_T + n_L$,
with
$n_T=A_Tf_T(R/R_T)$ and
$n_L=A_Lf_L(R/R_L)$. We then define $f_T$ and
$f_L$ to be normalised so that
$\int_0^{\infty}f_T(Z_T)dZ_T = \int_0^{\infty}f_L(Z_L)dZ_L = 1$,
where
$Z_T=R/R_T$ and $Z_L=R/R_L$. Conservation of total number of
droplets then implies
\begin{equation} A_TR_T + A_LR_L =n_0 \label{numcon}
\end{equation} Defining $\int_0^{\infty}f_T Z_T^3dZ_T/C_{\infty}v_b
= B_{0T}$ and
$\int_0^{\infty}f_L Z_L^3dZ_L/C_{\infty}v_b = B_{0L}$,
then
conservation of volume fraction implies
\begin{equation}
\frac{4\pi}{3}(A_TR_T^4(\epsilon)B_{0T} + A_LR_L^4(t)B_{0L}) +
\epsilon = \epsilon_0
\end{equation}
or
\begin{equation} A_LR_L(t) = \frac{(\epsilon_0-\epsilon) -
B_{0T}A_{T}(4{\pi}/3)R_T(\epsilon)^4} {(4{\pi}/3)B_{0L}R_L(t)^3}
\end{equation}
Substitution into Eq.\ref{numcon}  for $A_T$ then gives
\begin{equation} A_T = \frac{n_0}{R_T(\epsilon)} -
\frac{(\epsilon_0 - \epsilon )  -
B_{0T}A_{T}(4{\pi}/3)R_T(\epsilon)^4}
{(4{\pi}/3)B_{0L}R_L(t)^3R_T(\epsilon)}
\end{equation} which, to the order required, is Eq.\ref{at}.
Likewise for $A_L$ we obtain
\begin{equation} A_L = \frac{(\epsilon_0 -
\epsilon ) - n_0 B_{0T}(4{\pi}/3)
R_T(\epsilon)^3}{(4{\pi}/3)B_{0L}R_L(t)^4}
\end{equation} Substituting Eq.\ref{asymp} for $R_T(\epsilon)$ and
Eq.\ref{invert} for $\epsilon (R_L)$ yields
\begin{equation} A_L = \frac {\epsilon_0 -
n_0B_{0T}(4{\pi}/3)R_{B}^3} {(4{\pi}/3)B_{0L}R_L(t)^4} - \frac {(1 +
(3R_B/2)B_{0T}(4{\pi}/3)R_{B}^3) } {(4{\pi}/3)B_{0L}R_L(t)^5}
\end{equation}
Since Eq.\ref{upert} determines that 
droplets in the neighbourhood of $R_B$ will relax at an exponential
rate towards $R_T$, $f_T$ (which is
normalized to unity) will converge at an exponential rate onto  
$f_T={\delta}(Z_T-1)$. 
Hence $B_{0T}$ equals $1/(C({\infty})v_b)$ and $A_L$ becomes
\begin{equation} A_L = \frac {\epsilon_0 -
(n_0(4{\pi}/3)R_{B}^3)/(C({\infty})v_b)} {(4{\pi}/3)B_{0L}R_L(t)^4}
- \frac {(1 + (3R_B/2)(4{\pi}/3)R_{B}^3)/(C({\infty})v_b) }
{(4{\pi}/3)B_{0L}R_L(t)^5}
\end{equation} which to the required order is Eq.\ref{ampl}.

\section{Determination of $n_L$}
\label{nl}
We have assumed that $n_L(R,t)$ satisfies (by itself) the continuity
equation in droplet size space,
\begin{equation}
\frac{{\partial}n_L(R,t)}{{\partial}t} = -
\frac{{\partial}(n_L(R,t)U(R,\epsilon))}{{\partial}R}
\end{equation}
Writing $n_L(R,t)=A_Lf_L(Z_L)$ we obtain
\begin{equation}
\frac{{\partial}(A_Lf_L(Z_L))}{{\partial}t} =
- \frac{\dot{R_L}}{R_L}Z_L A_L
\frac{{\partial}(f_L(Z_L))}{{\partial}Z_L} +
\frac{{\partial}A_L}{{\partial}t} f_L(Z_L)
\end{equation}
Defining
\begin{equation} K_1=\frac{{\epsilon}_0 -
n_0(4{\pi}/3)R_{B}^3/C({\infty})v_b}{(4{\pi}/3)B_{0L}}
\end{equation} and
\begin{equation}
K_2=\frac{1+(3R_B/2)B_{0T}(4{\pi}/3)R_{B}^3/C({\infty})v_b}
{(4{\pi}/3)B_{0L}}
\end{equation}  then we find
\begin{equation}
\frac{{\partial}A_L}{{\partial}t} = -4 \frac{\dot{R_L}}{R_L} A_L -
K_2\frac{\dot{R_L}}{R_L^6}
\end{equation} So we obtain
\begin{equation}
\frac{{\partial}n_L(R,t)}{{\partial}t} = - A_L\frac{\dot{R_L}}{R_L}
\left(4f_L +Z_L\frac{{\partial}(f_L(Z_L))}{{\partial}Z_L}\right) -
K_2\frac{\dot{R_L}}{R_L^6} f_L
\end{equation} We next write ${ {\partial} ( n_L(R,t) U(
R,{\epsilon} ) ) / { \partial } R }$ in terms of $Z_L$ to obtain
\begin{equation}
\begin{array}{ll}
\dot{R_L}R_L^2(4f_L + Z_L\frac{{\partial}(f_L(Z_L))}{{\partial}Z_L})
+
\frac{K_2\dot{R_L}}{R_L^3A_L}f_L =&
\left[ f_L\left(-\frac{1}{Z_L^2} +\frac{2}{Z_L^3}
-\frac{4(R_{\alpha}/R_L)^2}{Z_L^5} \right) \right. \\ &+
\left. \frac{{\partial}(f_L(Z_L))}{{\partial}Z_L}
\left( \frac{1}{Z_L} -\frac{1}{Z_L^2} +
\frac{(R_{\alpha}/R_L)^2}{Z_L^4}
\right) \right]
\end{array}
\end{equation} Expanding $1/(R_L^3A_L)$ to lowest order in
$1/R_L$ we obtain
$K_2\dot{R_L}/R_L^3A_L \rightarrow K_2\dot{R_L}R_L$; keeping
only the leading terms on the right hand side we are left with
\begin{equation}
\begin{array}{ll}
\dot{R_L}R_L^2(4f_L + Z_L\frac{{\partial}(f_L(Z_L))}{{\partial}Z_L})
+ K_2\dot{R_L}R_Lf_L =&
\left[ f_L\left(-\frac{1}{Z_L^2} +\frac{2}{Z_L^3}
\right) \right.
\\ & \left. + \frac{{\partial}(f_L(Z_L))}{{\partial}Z_L}
\left(\frac{1}{Z_L} -\frac{1}{Z_L^2}
\right) \right]
\end{array}
\end{equation} Since the R.H.S. is a constant, then as $t
\rightarrow \infty$  the L.H.S. must also approach a constant value.
Trying $R_L={\Gamma}t^{\beta}$ with
$\Gamma$ a constant, then
\begin{equation}
\dot{R_L}R_L^2=\beta\Gamma^3t^{3\beta-1}
\end{equation} and
\begin{equation}
\dot{R_L}R_L=\beta\Gamma^2t^{2\beta-1}
\end{equation} Clearly both will tend to zero if $\beta < 1/3$,
which would not correspond to a coarsening state. However if
$\beta > 1/3$ then the L.H.S of the equation would diverge at late
times. Hence $\beta=1/3$, and as $t \rightarrow
\infty$, $\dot{R_L}R_L \rightarrow 0$. Writing $\gamma = \Gamma^3/3$
then results in the  same equation as that obtained and solved by
Bray\cite{17}, namely Eq.\ref{brayeq}.


\begin{thebibliography}{99}
\bibitem{phifoot} This assumes that the volume occupied by
a given amount of trapped species is independent of its concentration.
\bibitem{footsurprise}
This is perhaps surprising, since according to Eq.\ref{asymp},
the decaying supersaturation
$\epsilon(t)$ causes the small droplets to give up material at a
rate that could perturb $\dot\epsilon$ significantly; this might be
expected to lead to a slower decay of $\epsilon(t)$ and hence (via
Eq.\ref{invert}) of $R_L(t)$. However, this is not a correct
argument: for, as mentioned above, the Lifshitz-Slyozov mechanism
leads to an autonomous equation for $R_L(t)$ whose time development
is then entirely fixed by the self-consistency requirement on
$\gamma$.
Given the behaviour of $R_L(t)$, the ambient
supersaturation
$\epsilon(t)$ follows directly from
Eq.\ref{invert}; the latter in turn dictates the
evolution of the size $R_T(t)$ of the small droplets, via
Eq.\ref{rootl}.
\bibitem{footnote1}
Where $\overline{V}$ has now been defined to include the volume of the
bulk phase.
\end{thebibliography}
\end{document}